\documentclass[11pt,twoside]{article}
\usepackage{./asp2014}

\aspSuppressVolSlug
\resetcounters

\begin{document}

\title{Advanced Environment for Knowledge Discovery in the VIALACTEA Project}
\author{Ugo~Becciani,$^1$ Marilena~Bandieramonte,$^1$ Massimo~Brescia,$^2$ Robert~Butora,$^3$ Stefano~Cavuoti,$^2$ Alessandro~Costa,$^1$ Anna~Maria~di~Giorgio,$^4$ Davide Elia,$^4$ Akos~Hajnal,$^5$ Peter~Kacsuk,$^5$ Scige~John~Liu,$^4$ Amata~Mercurio,$^2$ Sergio~Molinari,$^4$ Marco~Molinaro,$^3$  Giuseppe~Riccio,$^2$ Eugenio Schisano,$^4$ Eva~Sciacca,$^1$ Riccardo~Smareglia,$^3$ Fabio~Vitello$^1$
\affil{$^1$ INAF Osservatorio Astrofisico di Catania, Catania, Italy; \email{ugo.becciani@oact.inaf.it}}
\affil{$^2$ INAF Osservatorio Astronomico di Capodimonte, Napoli, Italy; \email{brescia@oacn.inaf.it}}
\affil{$^3$ INAF Osservatorio Astronomico di Trieste, Trieste, Italy; \email{molinaro@oats.inaf.it}}
\affil{$^4$ INAF Istituto di Astrofisica e Planetologia Spaziali, Roma, Italy; \email{sergio.molinari@iaps.inaf.it}}
\affil{$^5$ SZTAKI Laboratory of Parallel and Distributed Systems, Budapest, Hungary; \email{kacsuk@sztaki.hu}}}

\paperauthor{Sample~Author1}{Author1Email@email.edu}{ORCID_Or_Blank}{Author1 Institution}{Author1 Department}{City}{State/Province}{Postal Code}{Country}
\paperauthor{Sample~Author2}{Author2Email@email.edu}{ORCID_Or_Blank}{Author2 Institution}{Author2 Department}{City}{State/Province}{Postal Code}{Country}
\paperauthor{Sample~Author3}{Author3Email@email.edu}{ORCID_Or_Blank}{Author3 Institution}{Author3 Department}{City}{State/Province}{Postal Code}{Country}

\begin{abstract}
The VIALACTEA project aims at building a predictive model of star formation in our galaxy. We present the innovative integrated framework and the main technologies and methodologies to reach this ambitious goal.
\end{abstract}

\section{Introduction}

The Milky Way galaxy is a complex ecosystem where a cyclical transformation process brings diffuse baryonic matter into dense unstable condensations to form stars, that produce radiant energy for billions of years before releasing chemically enriched material back into the InterStellar Medium in their final stages of evolution. Although considerable progress has been made in the last two decades in the understanding of the evolution of isolated dense molecular clumps toward the onset of gravitational collapse and the formation of stars and planetary systems, a lot remains still hidden.


The aim of the European FP7 VIALACTEA project is to exploit the combination of all new-generation surveys of the Galactic Plane to build and deliver a galaxy scale predictive model for star formation of the Milky Way.
The technological objectives of the project are: to boost the scientific exploitation of ESA missions space data by developing new and carefully tailored data processing tools 
to combine in a VO-compatible and interoperable way the new-generation Galactic Plane surveys;
to build and visualize an innovative 3D representation of the Milky Way Galaxy.

\section{VIALACTEA Technological Framework}

To explore very large regions of the galaxy, a new framework is implemented using advanced visual analytics techniques, data mining methodologies, machine learning paradigms and based on VO (Virtual Observatory) data representation and retrieval standards. All such specialized tools are integrated into a virtualized computing environment, resulting as an efficient and easy-to-use gateway for the scientific stakeholder community. An overview of the methodologies and technologies (see Figure \ref{vl_framework}), able to fulfil the scientific expectations of the project is presented in the following sections.


\articlefigure[width=.6\textwidth]{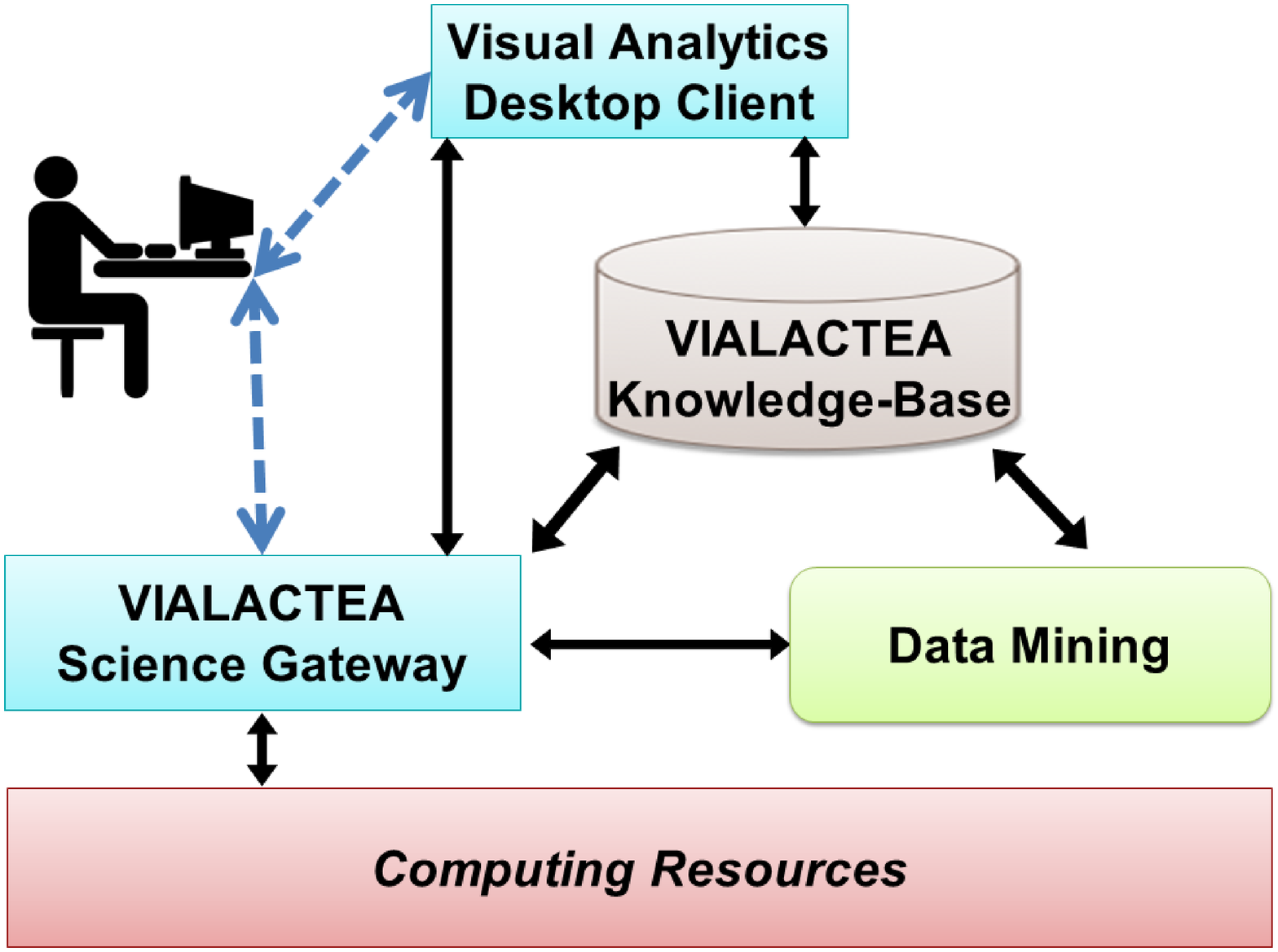}{vl_framework}{Schema of the interactions between the developed technological tools.}

\subsection{Database and Virtual Observatory Infrastructure}
\label{vlkb}

The ViaLactea Knowledge Base (VLKB) includes a combination of storage facilities, a Relational Data Base (RDB) server and web services on top of them.
The RDB is the content holder of the metadata of the stored files as well as a data resource itself. It is also the base resource on top of which the IVOA TAP service \citep{dowler2010table} works.
The goal of the VLKB RDB component is to allow easier searches and cross correlations between
VIALACTEA data using the software tools the user community have at its disposal. Consuming the
VLKB alongside other VO available resource is possible through the implementation of the TAP service
so that the project's community can exploit the VIALACTEA data without the need to continuously
retrieving and downloading external resources.

\articlefigure[width=.8\textwidth]{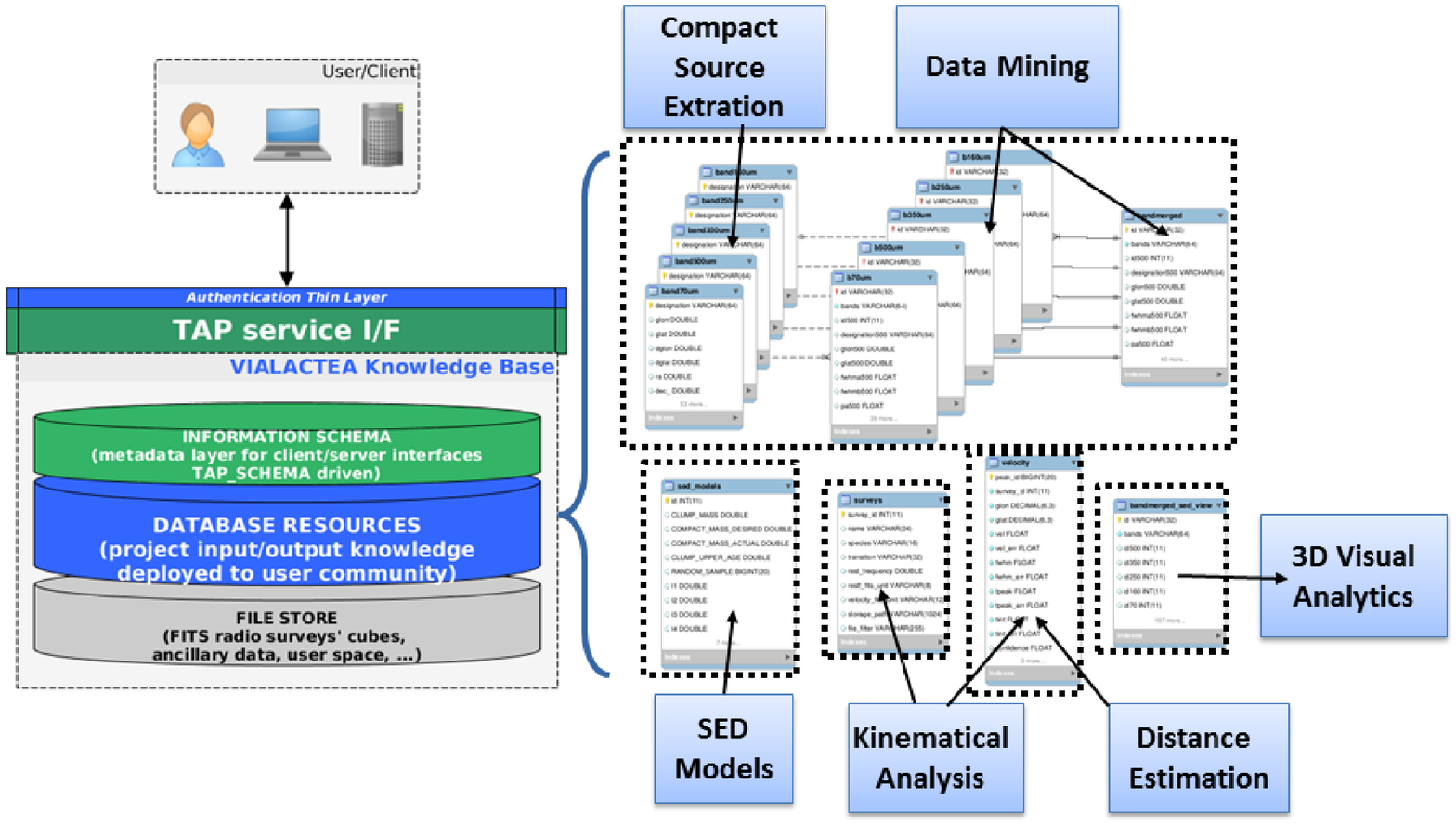}{vlkb_overview}{Overview of the VLKB schemas.}

Currently the RDB is subdivided, for logical and functional reasons, into separate schemas (see Figure \ref{vlkb_overview}) containing: Hi-GAL survey \citep{molinari2010hi} catalogue sources and related band merged information; structural informations such as filament structures or bubbles; and Radio Datacubes with search and cutout services. Finally a TAP\_SCHEMA is built for the VLKB TAP services, this allows to make all tables accessible using standard VO tools capable of TAP/ADQL queries.



\subsection{Data Mining Systems}
\label{DM}

The integration and exploitation of data-mining and machine-learning technologies within the project enable developing new tools that incorporate data products and the astronomer's know-how into a set of supervised workflows with decision-making capabilities to carry out building of Spectral Energy Distributions, distance estimate and Evolutionary classification of hundreds-of-thousands of star forming objects on the Galactic Plane.

More in detail, the main developed data mining tools are related to: compact source extraction to obtain a more refined version of band-merged catalogues based on the positional cross-match among sources at different wavelengths (Q-FULLTREE); filamentary structure detection to refine and optimize the detection of the edges of filamentary structures \citep{riccio2015machine} (FilExSeC); and source kinematical distance estimation combining all available information from Galactic rotation curve, spectroscopic survey data in molecular gas lines or 3D extinction maps in the near and mid-Infrared (MLNPQNA).

\subsection{3D Visual Analytics Systems}


Real-time data interaction are performed to carry out complex tasks for multi-criteria data/metadata queries on the VLKB, subsample selection and further analysis processed over the Science Gateway, or real-time control of data fitting to theoretical models.

\articlefiguretwo{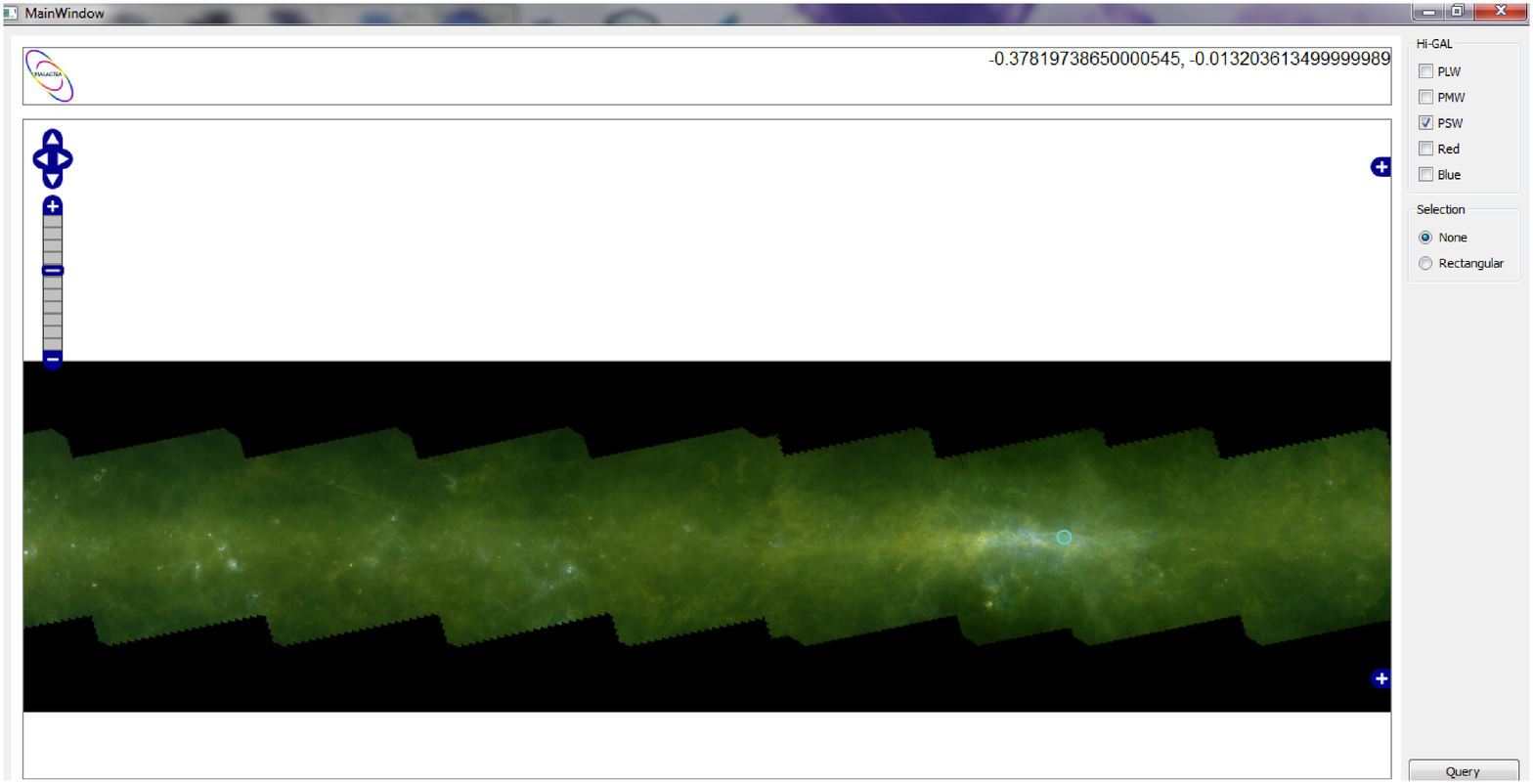}{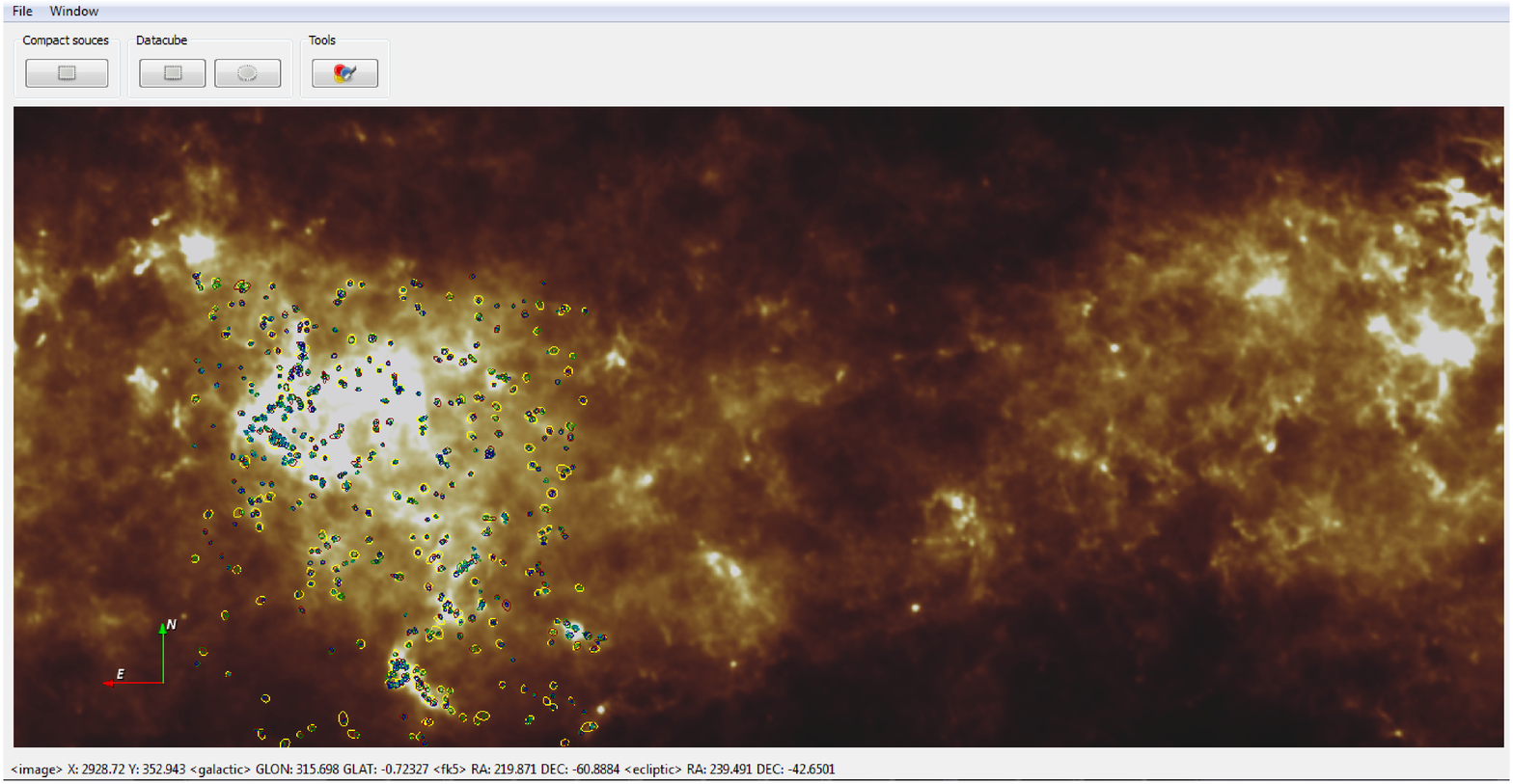}{visual_analyt}{Screenshots of the 3D Visual Analytics tool. \emph{Left:} Navigable overview of the galactic plane from -60 to +60 degree.  \emph{Right:} Compact sources visualization.}

The tool allows the user to query and download fits images from the search and cutout service provided from the VLKB infrastructure.
The navigation starting interactive interface is represented by the galactic plane mosaic from -60 to +60 degree (see Figure \ref{visual_analyt} on the left).
Starting from a selected region the relative fits image is downloaded from the VLKB and visualized. A query to the VLKB allows the scientist to obtain Hi-Gal bandmeged sources (as produced by Q-FULLTREE) that are shown on the map (see Figure \ref{visual_analyt} on the right). The user can select a sub-region of interest to study the contained SEDs. From each SED the fitting with theoretical models (stored in VLKB) and with analytical models may be performed. Finally the 3D Radio Datacubes visualization allows to analyze single slices and extract iso-contours to investigate the region of interest.

\subsection{Science Gateway}


The VIALACTEA Science Gateway\footnote{\url{http://via-lactea-sg00.iaps.inaf.it:8080/}} aims at serving as a central workbench for the VIALACTEA community. It is based on WS-PGRADE/gUSE \citep{kacsuk2012ws} portal framework which provides several ready-to-use functionalities off-the-shelf.



An infrastructure monitoring has been developed to guarantee full operation of the VIALACTEA infrastructures where a diversity of software and parallel and multi-thread jobs are running. The developed monitoring features of the
Science Gateway test the infrastructures health periodically, report monitoring tests on the gateway as well as the history of the monitoring tests and send e-mail alerts on failure.

\section{Conclusion}
The presented tools developed within the VIALACTEA project will be able to access data products as well as libraries of millions of radiative transfer models necessary for the science analysis in an integrated environment. Especially the emerging field of visual analytics brings data analysis and visualization into a single human-in-the-loop process. New technologies such as new 3D images of our galaxy with interactive data manipulation capabilities will provide powerful analytical, educational and inspirational tools for the next generation of researchers.

\acknowledgements The research leading to these results has received funding from the European Union Seventh Framework Programme (FP7/2007-2013) under grant agreement no. 607380 (VIALACTEA: the Milky Way as a star formation engine).

\bibliographystyle{asp2014}
\bibliography{P010}  

\end{document}